\documentclass[sigconf]{acmart}
\usepackage[nolist,nohyperlinks]{acronym}
\usepackage{subcaption}

\copyrightyear{2025}
\acmYear{2025}
\setcopyright{acmlicensed}\acmConference[RecSys '25]{Proceedings of the Nineteenth ACM Conference on Recommender Systems}{September 22--26, 2025}{Prague, Czech Republic}
\acmBooktitle{Proceedings of the Nineteenth ACM Conference on Recommender Systems (RecSys '25), September 22--26, 2025, Prague, Czech Republic}
\acmDOI{10.1145/3705328.3748030}
\acmISBN{979-8-4007-1364-4/2025/09}

\author{Bruno Sguerra}
\authornote{Contact author: \href{research@deezer.com}{research@deezer.com}}
\affiliation{
  \institution{Deezer Research}    
  \country{France}
}

\author{Elena V. Epure}
\affiliation{
  \institution{Deezer Research}
  \country{France}
}

\author{Harin Lee}
\affiliation{
  \institution{Max Planck Institute for Human Cognitive and Brain Sciences, Max Planck Institute for Empirical Aesthetics}
    \country{Germany}
}
\author{Manuel Moussallam}
\affiliation{
  \institution{Deezer Research}
    \country{France}
}

\begin{document}

\title{Biases in LLM-Generated Musical Taste Profiles for Recommendation}

\begin{abstract}
One particularly promising use case of Large Language Models (LLMs) for recommendation is the automatic generation of Natural Language (NL) user taste profiles from consumption data. These profiles offer interpretable and editable alternatives to opaque collaborative filtering representations, enabling greater transparency and user control. However, it remains unclear whether users consider these profiles to be an accurate representation of their taste, which is crucial for trust and usability. 
Moreover, because LLMs inherit societal and data-driven biases, profile quality may systematically vary across user and item characteristics. In this paper, we study this issue in the context of music streaming, where personalization is challenged by a large and culturally diverse catalog. We conduct a user study in which participants rate NL profiles generated from their own listening histories.
We analyze whether identification with the profiles is biased by user attributes (e.g., mainstreamness, taste diversity) and item features (e.g., genre, country of origin). We also compare these patterns to those observed when using the profiles in a downstream recommendation task. Our findings highlight both the potential and limitations of scrutable, LLM-based profiling in personalized systems.

\end{abstract}

\begin{acronym}
    \acro{MF}{Matrix Factorization}
    \acro{NDCG}{Normalized Discounted Cumulative Gain}
    \acro{CF}{Collaborative Filtering}
    \acro{SVD}{Singular Value Decomposition}
    \acro{LE}{Listening Event}
    \acro{LLMs}{Large Language Models}
    \acro{NL}{Natural Language}
    \acro{CI}{Confidence Interval}
    \acro{GS-Score}{Generalist-Specialist Score}
    \acro{Gold-MSI}{Goldsmiths Musical Sophistication Index}
    \acro{DR}{Doubly Robust}
    \acro{ATE}{Average Treatment Effect}
    
\end{acronym}


\maketitle

\section{Introduction}
\label{sec:intro}
In 1967, the American writer John M. Culkin said ``we shape our tools and thereafter our tools shape us'', highlighting the feedback loops between human experience and technologies.
Today, this dynamic is particularly evident in the rise of \ac{LLMs}, whose remarkable capabilities in text generation have far-reaching implications beyond academia and industry. Culkin's feedback loop also underscores a critical challenge: LLMs inherit human biases from their training data, which can result in discriminatory outcomes, especially when using sensitive attributes like gender~\cite{dai2024bias}. As LLMs influence human decision-making and generate content that may be used for future training, there is a risk of reinforcing and amplifying existing biases over time~\cite{brinkmann2023machine}.

In recommendation, a popular use of LLMs is to generate \ac{NL} user profiles from consumption data (item metadata and possibly explicit feedback)~\cite{radlinski2022natural}.
These profiles can then guide the recommendation, either via direct LLM prompting~\cite{ramos2024transparent} or by embedding users and items in a shared space~\cite{gao2024end}. The resulting LLM textual profiles offer a scrutable alternative to latent user embeddings common in~\ac{CF}~\cite{radlinski2022natural, ramos2024transparent}, allowing users to inspect and edit their profiles, thereby enhancing transparency and control. This approach also benefits cold-start cases, whereby new users can simply describe their preferences~\cite{radlinski2022natural}. One might assume that, since LLMs are not primarily trained on user-item interactions, they might avoid data-driven biases such as popularity, exposure, and position bias, which are common in recommendation~\cite{matrosova2024recommender,ungruh2024putting,chen2023bias,huang2024going,pipergias2023collaborative}. However, they introduce new challenges such as hallucinations~\cite{jiang2024beyond}, sensitivity to item order~\cite{wang2023large}, and dependence on memorized content~\cite{carlini2022quantifying,di2025llms}. Moreover, recent findings show that LLMs as recommenders tend to exhibit lower fairness and diversity than traditional \ac{CF} methods, though techniques like fine-tuning and prompt design may help address these issues~\cite{deldjoo2024understanding,dai2024bias}.

Studies have shown that prompting LLMs with item metadata and feedback can yield coherent textual user summaries~\cite{ramos2024transparent, zhou2024language, gao2024end}, with evaluation often focusing on downstream performance. However, it remains unclear to what extent users perceive these profiles as accurate representations of themselves (i.e., their preferences). Moreover, whether (and how) profile representativeness and usability in recommendation vary with user attributes (e.g., mainstreamness) or item characteristics (e.g., music genre), particularly in combination with the social biases LLMs inherit from pre-training, remains an underexplored aspect.

In this work, we address this gap by evaluating profile quality along two dimensions: user-self identification, collected through a user study, and recommendation relevance, assessed in a downstream task. We argue that understanding how these dimensions relate to user and item characteristics is essential for assessing the feasibility of LLM-generated NL profiles, before optimizing them for downstream applications. In particular, such analysis is necessary to ensure that all user groups are fairly represented and accounted for. 

We address the following research questions: (1) To what extent users recognize their musical taste in automatically generated profiles? (2) Is profile quality biased towards user characteristics such as mainstreamness or consumption diversity, or item characteristics such as genre and country of origin? (3) Is user evaluation of profile quality aligned with relevance in recommendation settings?

We conduct our study in the context of music streaming, where modeling user preferences is especially challenging due to the scale and diversity of catalogs spanning genres, regions, and popularity levels. To support transparency and reproducibility, we release the source code of our experiments and the user study dataset (see Section~\ref{sec:downstream}).

\section{Experimental Setup}
A key lacking aspect of recent LLM summarization methods is human qualitative evaluation of their ability to correctly represent user preferences. Therefore, we conducted a user study to investigate potential biases in NL user profiles following two stages: (1) generating NL profiles in a systematic way to understand which features matter on the resulting quality; (2) conducting an online study (N participants = 64) to evaluate the generated profiles.



\subsection{Building NL User Musical Profiles}
\label{sec:BuildNLUser}

Generating scrutable \ac{NL} user profiles offers transparency but necessitates conciseness, a significant challenge in music recommendation. Unlike prior work often relying on explicit feedback~\cite{ramos2024transparent, zhou2024language, gao2024end}, music recommendation frequently relies on implicit feedback, which is more noisy, making preference inference more uncertain~\cite{jawaheer2014modeling, sguerra2025uncertainty}. While using concatenated item attributes from consumption logs can help model preferences in implicit settings~\cite{xu2024prompting}, in music consumption, users interact with vast catalogs~\cite{schedl2019deep}, leading metadata-based representations to exceed LLM context limits. This makes item sampling strategies particularly important. Furthermore, musical taste blends long-term preferences with short-term exploration~\cite{tran2024transformers}.
Thus, sampling only frequent tracks within short windows (e.g., one week) favors recency but is prone to noise. In contrast, longer time windows smooth out fluctuations but could miss more recent explorations. Also, since users listen to collections in the form of playlists and albums, simple frequency-based sampling can lead to redundant profiles due to repeated listens of the same collections. 

To address these challenges, we propose a 2-step sampling strategy:
(1) identify the top-n most played artists within a given time window; (2) select most played tracks per artist. This mitigates artist-level redundancy and allows the time window to control the balance between short- and long-term preference signals. 

\begin{figure}[htbp]
    \centering
    \includegraphics[width=0.79\linewidth]{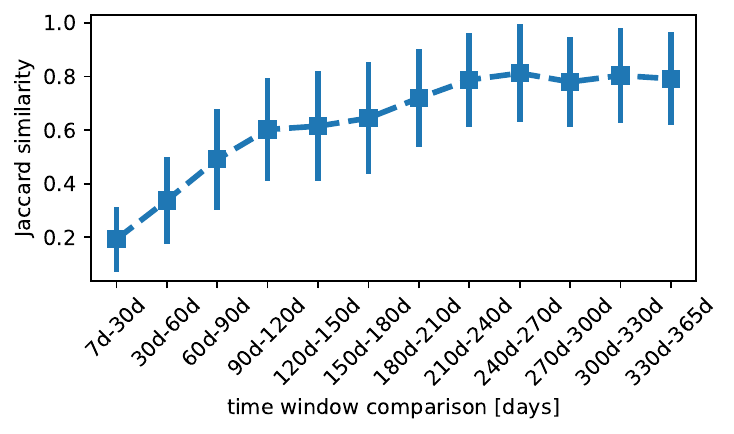}
    \caption{Mean and standard deviation of Jaccard similarity for users' top-15 tracks, across different time window lengths.}
    \label{fig:jaccard_window}
\end{figure}

Figure~\ref{fig:jaccard_window} illustrates the increasing stability of sampled top artist-track sets across different time windows. Each point represents the mean Jaccard similarity across users (with standard deviation bars), comparing top tracks from a shorter window to those from a longer one. For example, the first point compares users’ top tracks in a 7-day window (``7d'') to those in a 30-day window (``30d''). The upward trend indicates that as the time windows increase, the sampled top items become more stable. In our study, we generate profiles using our two-step strategy across four time windows: two shorter ones (30 and 90 days), where preferences are still relatively unstable, and two longer ones (180 and 365 days), where preferences tend to plateau.

For \ac{NL} summary generation, we follow prior work by prompting \ac{LLMs} with a list of items described by textual metadata. For each user, we sample their top-15 artist–track pairs using our two-step strategy and collect metadata
(track title, play count, artist name, album name, release year, artist main and secondary genres). The number of items was chosen to fit within a single-shot prompt, avoiding context length limitations. Prior work often uses fewer items (5 items in~\cite{gao2024end, ramos2024transparent}) and sometimes more (30 in~\cite{zhou2024language}). However, our results show 15 items are enough for user recognition.


We do not provide example profiles in the prompt, allowing the model to propose its own structure. This choice enables us to assess which formats resonate best with users. Also, the prompt explicitly instructs the model not to overly rely on artist and track names, encouraging more abstract and high-level descriptions of preferences. 
We select two open-source LLMs, Llama 3.2~\cite{touvron2023llama} and DeepSeek-R1~ \cite{guo2025deepseek} (a reasoning-focused model), both available via the Ollama library~\cite{ollama2025}, and Gemini 2.0 Flash~\cite{google2025gemini}, a proprietary model. We leave the output context window to their default values
and set the temperature to 0.8. We include the full prompt and profile examples in the Appendix.


\subsection{User Study}

We recruited participants on a voluntary basis through internal communication channels within our organization (Deezer). Participation took place during paid working hours, and the study was conducted online. Upon invitation, participants were explicitly informed that their streaming histories would be used to generate profiles for evaluation purposes.

A total of 64 participants took part in the study (40 males, 23 females, and one non-binary), aged between 22 and 48 years (33.73 $\pm$ 6.28). 
For them, we collected several high-level metrics, derived from 28-day consumption data on Deezer, a music streaming service: (1) \ac{GS-Score}\footnote{It quantifies the dispersion of songs users have listened to based on SVD embeddings derived from co-occurrence of tracks in playlists and personal collections.}, which captures whether a listener is a specialist (high GS-score) or a generalist (low GS-score); 
(2) the median rank of their listened tracks,  relates to the popularity of the tracks based on the consumption of all users of Deezer is as a measure of mainstreamness, higher values indicating a preference for popular tracks;
(3) mean age of songs
indicates a preference for recent vs. older music. 

In addition, we collected participants’ consumption history for the year starting April 1, 2024, considering only streams with a listening time greater than 30 seconds\footnote{A threshold commonly used in the industry to define a positive interaction.}. From this data, we sampled items using the artist-track frequency method described in Section~\ref{sec:BuildNLUser}, across four time windows, and retrieved the corresponding metadata\footnote{Metadata sourced from Music Story: \url{https://music-story.com/}.}. We then generated an \ac{NL} profile using the three selected LLMs. Therefore, every user has 12 profiles. 

The mean profile length (in characters) per model is: Gemini: 396.43 $\pm$ 70.62; DeepSeek: 543 $\pm$ 153.06; Llama: 638.16 $\pm$ 117.54. Gemini-based profiles being overall smaller than the others. 

During evaluation, each participant was presented with 17 profiles: 12 personalized and five random profiles from other users serving as negative baselines. The evaluation begins with an introductory screen presenting an example profile to familiarize participants with the task. Then, the 17 profiles are presented in randomized order. For each profile, users are asked \textit{“How well does the following description match your music taste?”}, and asked to respond with a 7-point Likert scale (1 being "not at all" and 7 being "very much"). After rating all profiles, users are asked about their musical engagement based on \ac{Gold-MSI}~\cite{mullensiefen2014musicality}.

The user study was launched on April 9th 2025 and was concluded on the 11th. In the next section we discuss how the obtained user ratings relate to different user-item characteristics.

\begin{figure}[htbp]
    \centering
    \includegraphics[width=1\linewidth]{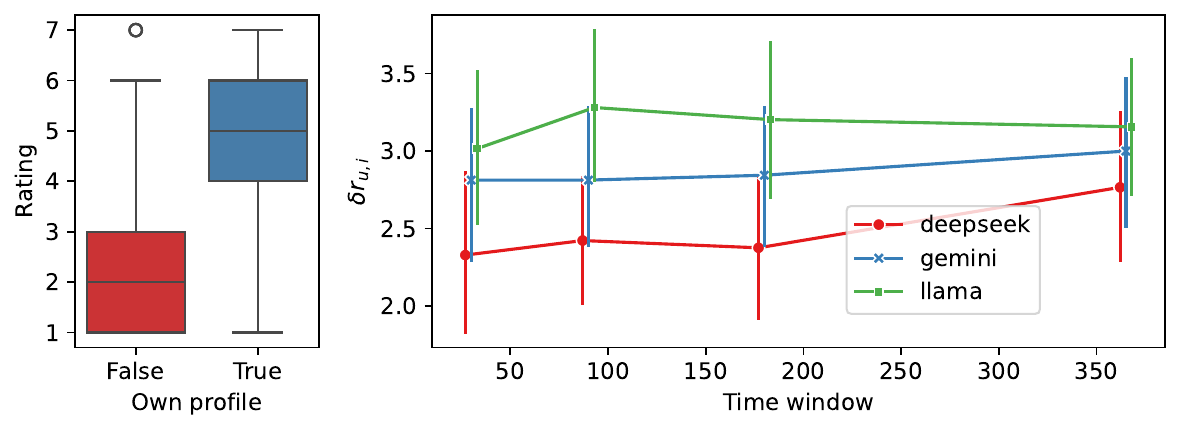}
    \caption{Left: box-plot of the user ratings for the true and false profiles. Right: the mean and 0.95 \ac{CI} for the profiles generated by each model at each time window.}
    \label{fig:ratings}
\end{figure}

\section{Biases in User-evaluated NL Profiles}

To investigate how well the generated profiles represent users in relation to different user- and item- characteristics, for every user $u$, we compute the difference in rating between a true profile $p_i \in {P}^+_u$ and the median rating on their fake profiles ${P}^-_u$ following: $\delta r_{u,i} = r_{u,i} - \text{median}\left(\{r_{u,j} : p_j \in \mathcal{P}^-_{u}\}\right)$, where $r_{u,i}$ is the rating user $u$ gave to profile $p_i$. Therefore, instead of using absolute ratings, that vary depending on users, we use this difference as a measure of how well users believe a profile represents their taste in relation to the random baseline. 

The left side of Figure~\ref{fig:ratings} shows box plots of ratings for positive and negative profiles, indicating that users rated profiles based on their own consumption data higher than random ones. On the right, we report the mean~$\delta r_{u,i}$ and 95\%~\ac{CI} for each model across four time windows. Time window had only a marginal impact on profile ratings, and although confidence intervals overlap, Llama-based profiles consistently received higher scores than the other two. This suggests that, beyond content, the form, specific to each LLM, may play a greater role in self-recognition. Supporting this, a two-way ANOVA revealed a significant effect of model, \textit{F}(2, 1076) = 5.16, \textit{p} = 0.0059, but no effect of time window, \textit{F}(3, 1076) = 1.59, \textit{p} = 0.191, and no interaction, \textit{F}(6, 1076) = 0.28, \textit{p} = 0.947.

\subsection{User-characteristics}
\label{sec:user-characteristics}
Here we evaluate how the ratings relate to user characteristics that are not directly dependent on the tracks used for the profiles. Our goal is to assess whether the method exhibits biases towards specific user groups characterized on some dimension, therefore we investigate linear relationships.

Concerning gender, rating differences were small across all models. Female users gave slightly lower ratings than male users, with Cohen’s $d$ indicating a small effect for DeepSeek ($d = -0.15$) and Gemini ($d = -0.16$), and a negligible effect for Llama ($d = -0.07$).

To quantify the influence of other user characteristics on rating behavior, we perform 5K bootstrap simulations, where we fit linear regression models predicting $\delta r_{u,i}$ from each investigated characteristic (normalized) for each LLM separately. This process allows us to quantify the~\ac{CI} for the fitted linear coefficient. The orange bars on the top of Figure~\ref{fig:linear_coeff} show the mean coefficient with the black bar indicating the 95\% CI. GS-score shows a significant positive association with $\delta r_{u,i}$, indicating that specialist users tend to rate the profiles more positively. This is consistent with the idea that specialist users, having narrower preferences, are more easily represented through the 15 sampled items, while generalist users may require more items for accurate representation. The mean age of consumed songs and user age also correlate positively with $\delta r_{u,i}$ across all models, suggesting that older users and users that listen to older songs tend to rate the generated profiles more favorably (these measures are both somewhat correlated Pearson = 0.22, p-value=0.048). User mainstreamness, measured via the median track rank, was not a significant predictor in any model (all \ac{CI} go through 0), suggesting no clear relationship between mainstream taste and perceived profile quality, the same for active engagement. 

\subsection{Item-characteristics}
\label{sec:items}
We next investigate how sampled item characteristics relate to the quality of profiles rated by users, using the same bootstrapping approach described above.

We first examined whether the median play count of the sampled items relate to user ratings, Deepseek profiles showed a positive association ($mean(\beta) = 2.11~[0.22, 3.93]$), suggesting repeated interactions may enhance relevance. Llama profiles showed a slightly weaker effect ($mean(\beta) = 1.91~[0.31, 3.55]$), while no clear effect was observed for Gemini ($mean(\beta) = 0.83~[-0.92, 2.78]$).

Since each track is annotated with a single ``main genre'' (when available), we compute, for each profile, the proportion of tracks associated with each genre among the 15 sampled tracks. 
We also include analogous ratios based on country of origin, as well as the proportion of missing genre or country annotations. We compute entropy measures over these distributions. Together, these features enable us to investigate how the compositional profile characteristics relate to their perceived quality. Other than the entropy measures, in the following analysis, we consider only genres and countries that appear in the listening histories of at least 20\% of the users. The set of investigated genres are: \textit{electronic, jazz, metal, pop, rap, rock, soul, world music}. The set of investigated countries: \textit{Belgium, Brazil, Canada, France, Germany, United Kingdom, United States}. 

\begin{figure}[htbp]
    \centering

    \begin{subfigure}[b]{\linewidth}
        \centering
        \includegraphics[width=0.95\linewidth]
        {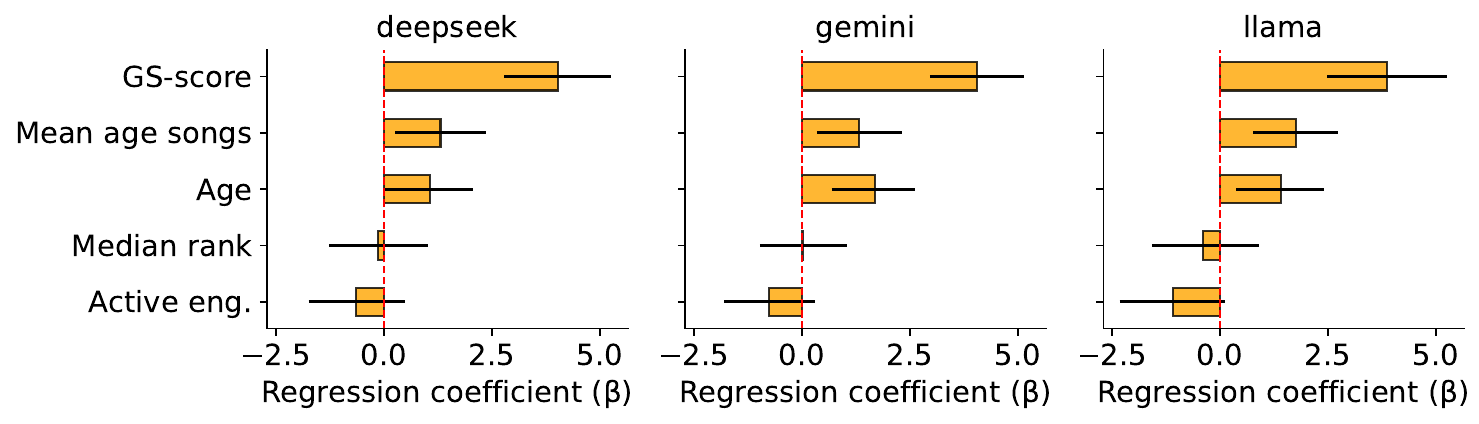}
    \end{subfigure}
    
    \begin{subfigure}[b]{\linewidth}
        \centering
        \includegraphics[width=0.95\linewidth]
        {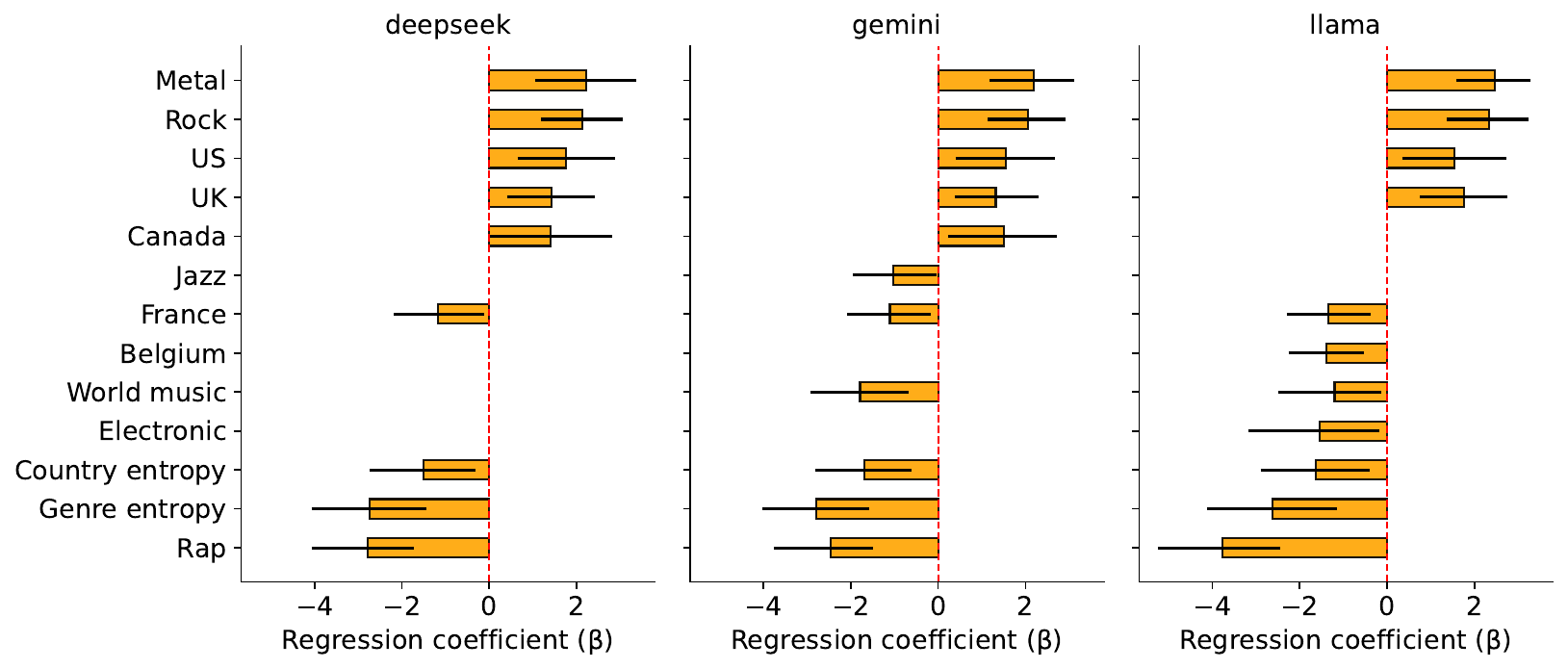}
    \end{subfigure}

    \caption{Bootstrapped estimated coefficients of user characteristics and genre and country composition on profile ratings.}
    \label{fig:linear_coeff}
\end{figure}

We perform 5K bootstrap simulations to estimate the linear association between these ratios and $\delta_{u,i}$. Figure~\ref{fig:linear_coeff} shows the bootstrapped mean coefficients and 95\% confidence intervals. We exclude associations with intervals crossing zero or with width greater than 10, as these indicate high uncertainty and lack of a reliable effect. The results suggest that specific item characteristics are consistently linked to profile quality, for example, a higher proportion of rap tracks is negatively associated with ratings across all models, while a greater share of U.S.-origin tracks correlates positively. 

\subsection{Biases in NL profiles}

While the linear coefficients shown above suggest that LLMs generate some profiles better than others over certain item features, this may reflect confounding factors, such as user preferences, rather than true model biases. For instance, profiles dominated by genres users already prefer (e.g., metal for a metal fan) may receive higher ratings simply due to taste alignment. In this case, being a metal fan acts as a confounder, influencing both the sampled items used to generate the profile and the user’s evaluation of it. As a result, the observed correlation between genre ratio and rating may be spurious with respect to assessing the LLM’s ability to generalize preferences. Additionally, users with a strong genre presence may be easier to summarize, particularly for specialists (top of Figure~\ref{fig:linear_coeff}).

To disentangle these effects, we apply a~\ac{DR} estimation of the \ac{ATE}~\cite{funk2011doubly}. This method combines outcome regression and propensity score modeling, yielding consistent estimates if either model is correctly specified. For a binary treatment variable $T$, 
the ATE is: $ATE = \frac{1}{N} \sum (\frac{T_i(Y_i - \hat{Y_1}(X))}{PS(X)} + \hat{Y_1}(X)) -  
    \frac{1}{N} \sum (\frac{(1 - T_i)(Y_i - \hat{Y_0}(X))}{1 - PS(X)} + \hat{Y_0}(X))$, 
where, $X$ is the covariates, $Y$ is the outcome (in our case $\delta r_{u,i}$), $\hat{Y}_1$ and $\hat{Y}_0$ are predicted outcomes given $X$ (via gradient boosting regression) under treatment and control, and $PS(X)$ is the propensity score from logistic regression.

\begin{figure}[htbp]
    \centering
    \includegraphics[width=1\linewidth]{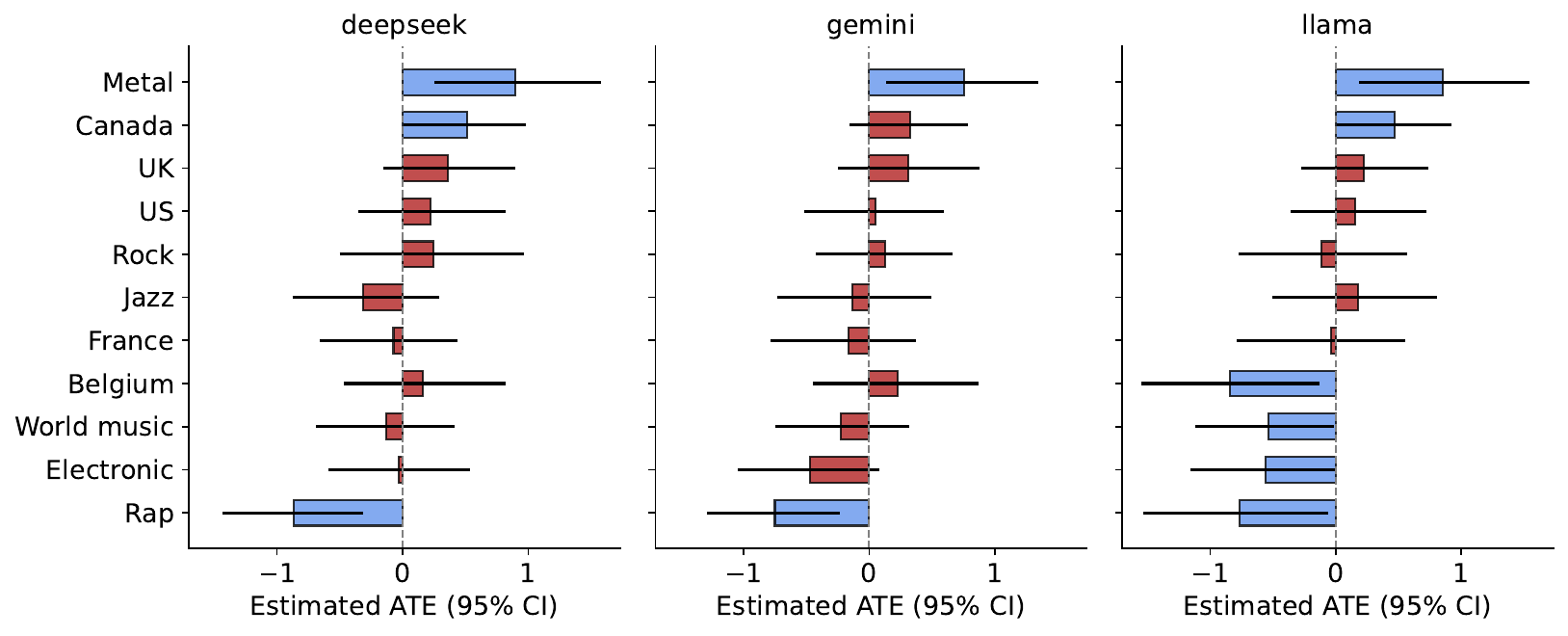}
    \caption{The estimated \ac{ATE} of the presence of a genre in the profiles to the ratings, controlled by the covariates. Blue bars show a consistent effect, while red bars correspond to estimates whose 95\% CI bar crosses zero, suggesting no significant effect.}
    \label{fig:bootstrapped_ATE}
\end{figure}

To control for taste and exposure biases, we construct covariates $X$ from each user’s long-term genre ratios, the genre composition of profile items (to capture underrepresentation effects), user age, and GS-score. We restrict the analysis to the genres shown in Figure~\ref{fig:linear_coeff} to reduce collinearity from rare categories like country or blues. Treatment is defined as having a genre ratio above the median, and the ATE quantifies the effect of including more content from this genre on user ratings, while controlling for confounders. For each genre and model, we estimate ATE and 95\% CI via 1K bootstrap simulations (Figure~\ref{fig:bootstrapped_ATE}).

The bootstrapped \ac{ATE} estimates in Figure~\ref{fig:bootstrapped_ATE} show that, even after controlling for our covariates, the presence of certain content types in the generated profiles still significantly influences user ratings. This is especially evident for rap and metal: across all models, rap-related content is associated with consistently negative ATEs, while metal shows positive ones. This suggests that profiles with a higher proportion of rap tracks tend to receive lower ratings. Sensitivity to content types also varies by model: for example, Gemini and DeepSeek are largely unaffected by the presence of Belgian and world music, while Llama see drops in profile quality. Also both DeepSeek and Llama see a positive effect from Canadian music that is not reliable for Gemini (CI crosses zero).

\section{Downstream Evaluation}
\label{sec:downstream}
We assess whether profile ratings correlate with downstream recommendation performance. Investigating if higher-rated \ac{NL} profiles also yield better recommendations, essential for ensuring both transparency and utility. We highlight that the goal is not to optimize recommendation performance, but rather to analyze its relation with user evaluations. 

Following prior work, we adopt a recommendation approach by embedding \ac{NL} profiles and item metadata into a shared latent space \cite{gao2024end, radlinski2022natural}.
Profiles are used as-is, while each track has associated metadata textual representation, as used in the prompts.
Given both items and users textual representations, we fine-tune the bi-encoder proposed in \cite{Epure2024Harnessing} on our dataset by using their music-specific ranker as a teacher\footnote{The ranker is implemented as a cross-encoders that take as input
a music description concatenated with a set of music tags and produces as output a similarity score.}.

We use all the generated profiles for training. 
For each profile, the sampled tracks for each user–time window pair serve as positive examples, while negative examples are sampled from tracks the user has not interacted with in the year considered.
In total, the training set comprised 100K user-positive-negative triplets.
We train the model for 1 epoch using an AdamW optimizer, a batch size of 10, and a maximum sequence length fixed to 512. 
To test the model, for each user, we sampled 10 items the user has interacted with (but not used for profile generation) and 1K negative items, all of which not seen in the training set with that specific user.
Our analysis is based on recall@10 and ndcg@10 metrics from the model's similarity scores, averaged over 5 runs with different random seeds. The code for implementation is made public in our repository\footnote{\href{https://github.com/deezer/recsys25_llm_biases}{https://github.com/deezer/recsys25\_llm\_biases}}.

\begin{figure}[ht]
    \centering
    \includegraphics[width=1\linewidth]{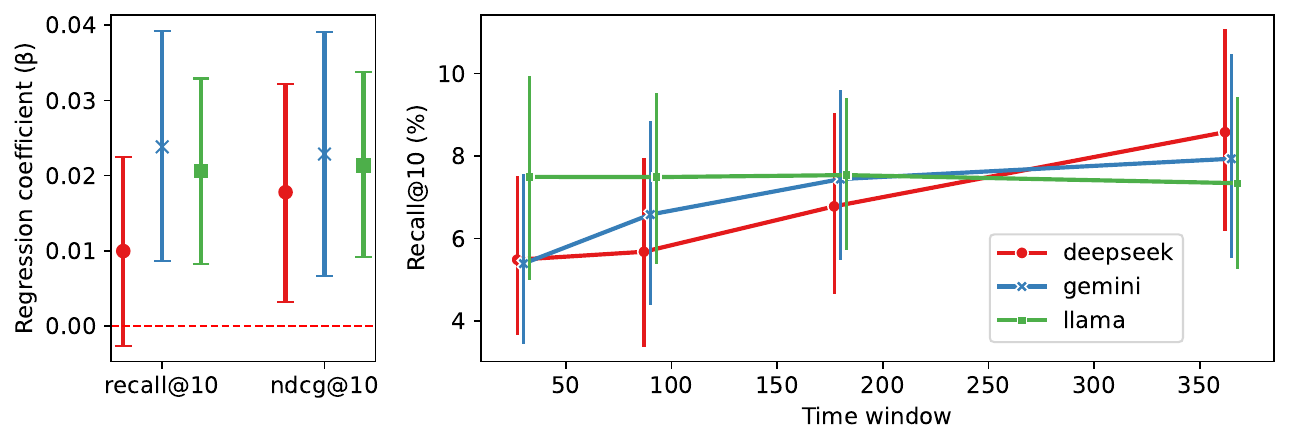}
    \caption{Left: Bootstrapped linear coefficients predicting recommendation scores from profile quality. Right: Mean Recall@10 with 95\% CI for each model across time windows.}
    \label{fig:recall}
\end{figure}

The left side of Figure~\ref{fig:recall} illustrates how $\delta r_{u,i}$ relates to recall@10 and ndcg@10, obtained by bootstrapping linear regression models that predict the normalized metrics from the user ratings, representing how much recommendation performance improves, on average, with a one-unit increase in user ratings. Except for Deepseek, all other metrics and models show a weak positive linear relationship. We provide more details on how recommendation performance relates to user and item characteristics in the Appendix. On the right side of Figure~\ref{fig:recall}, we show the mean recall@10 with 95\% CI for the different models and time windows. We note that the pattern is quite different from that in Figure~\ref{fig:ratings}. Except for Llama, the recommendation performance seems to improve with the longer time window, which can partially be explained by the fact that the test items were sampled from the entire consumption year.

\section{Discussion and Conclusion}

This paper assessed the quality of LLM-generated natural language user profiles along two dimensions: user ratings and downstream recommendation performance. We compared profiles generated from items sampled over varying time windows of users' listening histories and found that the time window had only a minor impact on ratings. As long as key preferences were captured, users generally recognized themselves in the profiles. While some noted missing aspects of their taste (see Appendix), the largest variation in ratings stemmed from the choice of LLM rather than the data window. This variation likely reflects stylistic differences in the generated profiles: Gemini, for instance, produced shorter, high-level descriptions; Llama often emphasized specific artists and track names (despite prompt instructions), which some users preferred. DeepSeek occasionally highlighted metadata such as ``remastered'' in track titles, which users found less meaningful. Yet given the diversity in user preferences and model-specific biases, it would require a deeper qualitative analysis to expose what specific aspects were preferred.

We further examined how user and item characteristics influenced profile quality. Profiles of specialist users (those with high GS-scores) tended to receive higher ratings, suggesting their tastes may be easier to capture in natural language. Item characteristics, such as genre and country ratios, were also correlated with ratings, both positively and negatively. To account for potential confounds such as taste alignment, we employed a \ac{DR} framework. The resulting ATE values confirmed that some content types systematically influenced ratings: for instance, rap content tended to reduce scores, while metal increased them, within the limits of our small user sample. Although it remains unclear whether these effects stem from metadata quality or model behavior, the fact that different LLMs responded differently to the same metadata suggests model-specific biases in content representation. These findings raise fairness concerns, as some users may consistently receive more representative profiles than others. Given that LLMs can inherit biases from the textual data on which they are trained, communities that are underrepresented in digital corpora—particularly non-English-speaking ones, may be disadvantaged. We note, however, that our study included only three LLMs, and conclusions may not generalize broadly.

Moreover, user ratings and recommendation performance were only weakly correlated, revealing a misalignment between perceived representativeness and algorithmic utility. This is critical: if users do not recognize themselves in recommendation-optimized profiles, they may lose trust in the system. One potential solution is to decouple profile generation for user-facing experiences from those optimized solely for recommendation.

Future work should explore targeted fine-tuning or debiasing strategies to reduce representation gaps. In production systems, LLMs could be refined using feedback from user-corrected profiles, aiming to better balance personalization and fairness.

\bibliographystyle{ACM-Reference-Format}

\appendix
\section*{Appendix}
\section{Complementary Data Analysis}
This appendix provides supplementary data explorations that complement the main paper.  

\begin{figure}[htbp]
    \centering
    \includegraphics[width=1\linewidth]{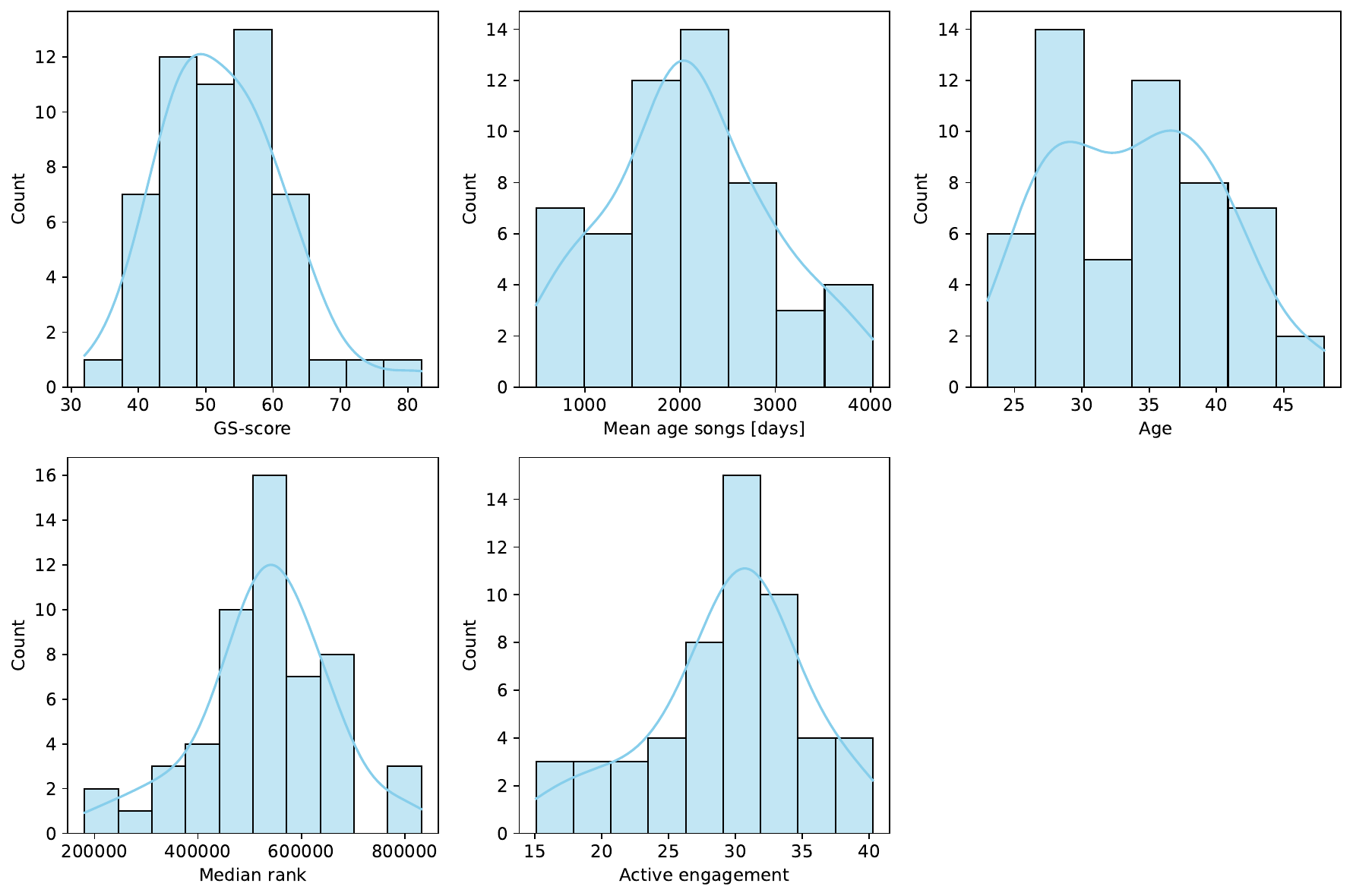}
    \caption{Distribution of the user characteristics.}
    \label{fig:user_characteristics_app}
\end{figure}

\begin{figure}[th]
    \centering
    \includegraphics[width=1\linewidth]{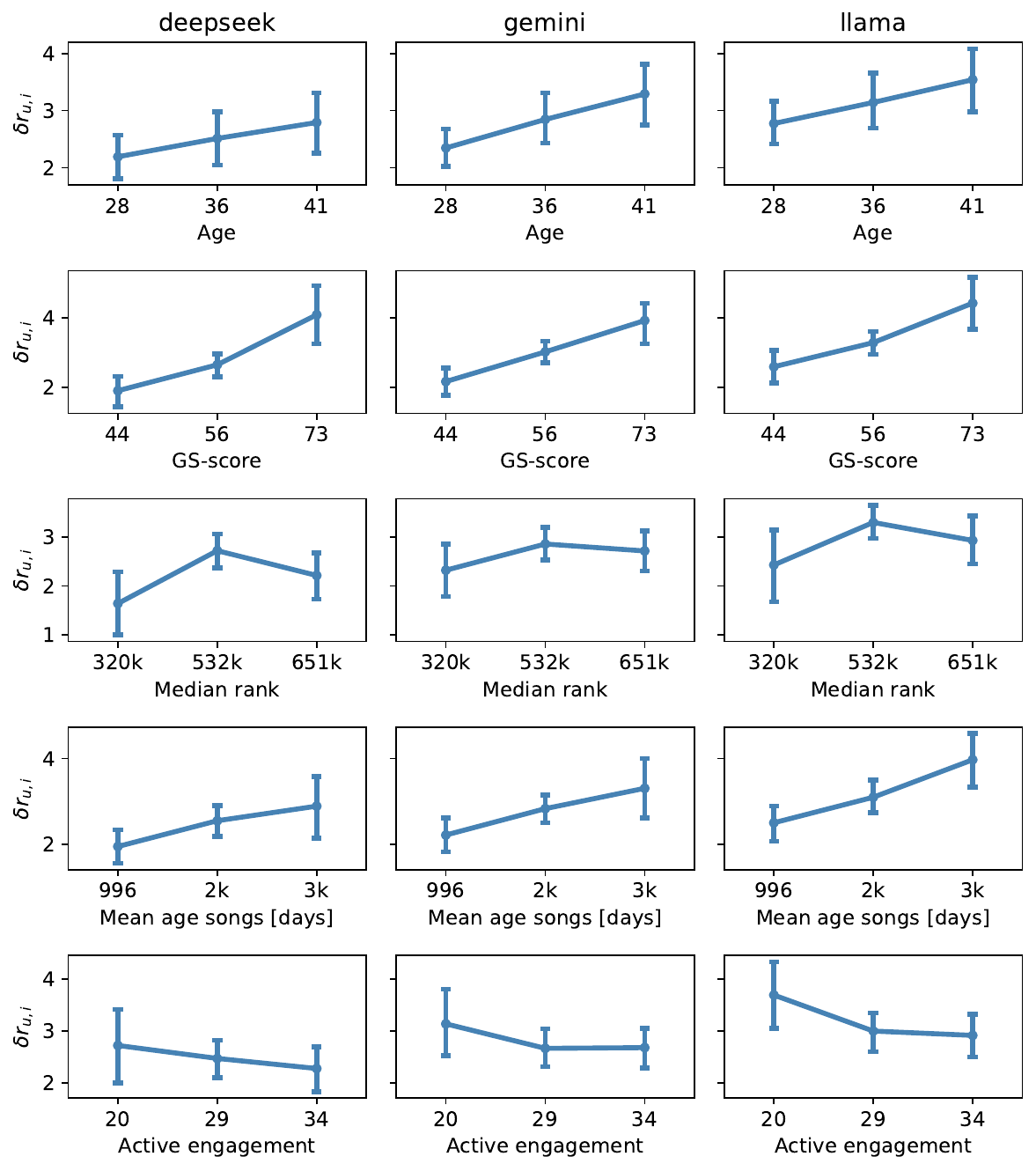}
    \caption{Mean $\delta r_{u,i}$ and 95\% \ac{CI} for user characteristics linearly binned into 3 groups.}
    \label{fig:linear_bins}
\end{figure}

Figure~\ref{fig:user_characteristics_app} shows the distribution of the considered users characteristics of the participants. Figure~\ref{fig:linear_bins} complement the explorations of Section~\ref{sec:user-characteristics}. Here, the evaluated characteristics are binned into 3 linear bins and we plot the mean $\delta r_{u,i}$ with a 95\% \ac{CI} for each bin. This serves for illustrating the relationship of the user characteristics and the ratings. 

\begin{figure}
    \centering
    \includegraphics[width=1\linewidth]{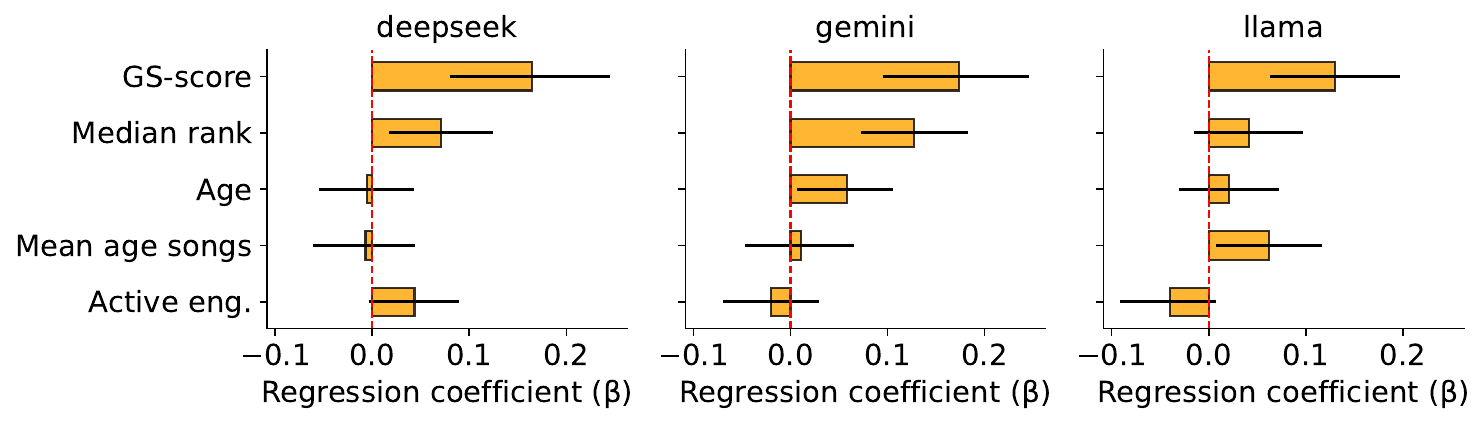}
    \caption{Bootstrapped estimated linear coefficients of user characteristics predicting recall@10.}
    \label{fig:user_recall}
\end{figure}

\begin{figure}
    \centering
    \includegraphics[width=1\linewidth]{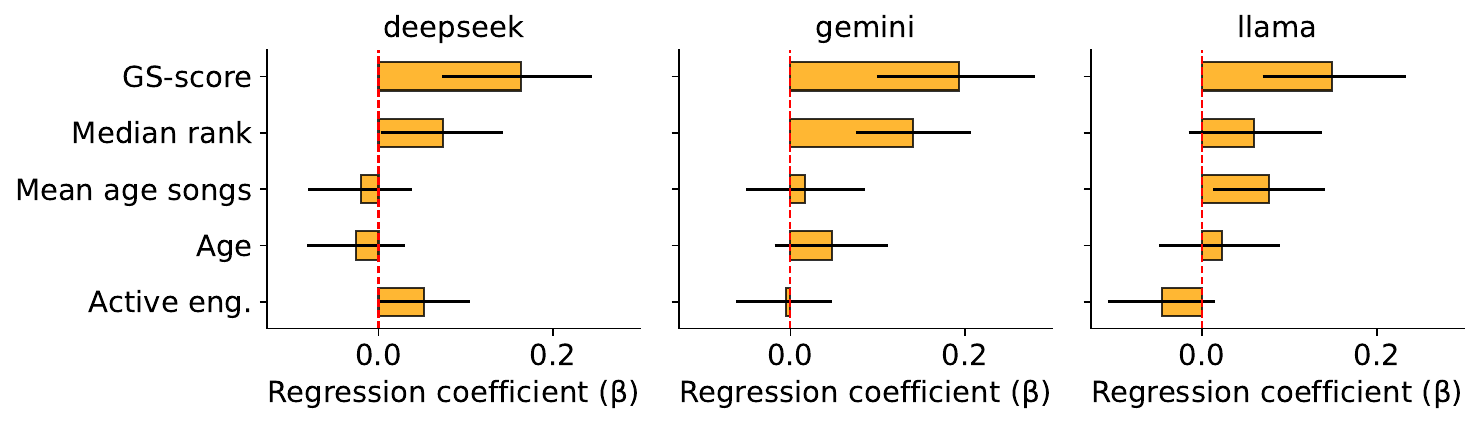}
    \caption{Bootstrapped estimated linear coefficients of user characteristics predicting ndcg@10.}
    \label{fig:user_ndcg}
\end{figure}

Figure~\ref{fig:user_recall} and Figure~\ref{fig:user_ndcg} illustrate the bootstrapped mean $\beta$ and 95\% CI from fitting 5K linear regression predicting Recall@10 and NDCG@10 from each considered user characteristics (normalized). Here we note that, contrary to the Figure~\ref{fig:linear_coeff}, the median rank, or the popularity of the tracks listened by the user, correlated positively with the metrics, meaning that users with preference for popular tracks have higher recommendation scores. 

\begin{figure*}
    \centering
    \includegraphics[width=0.8\linewidth]{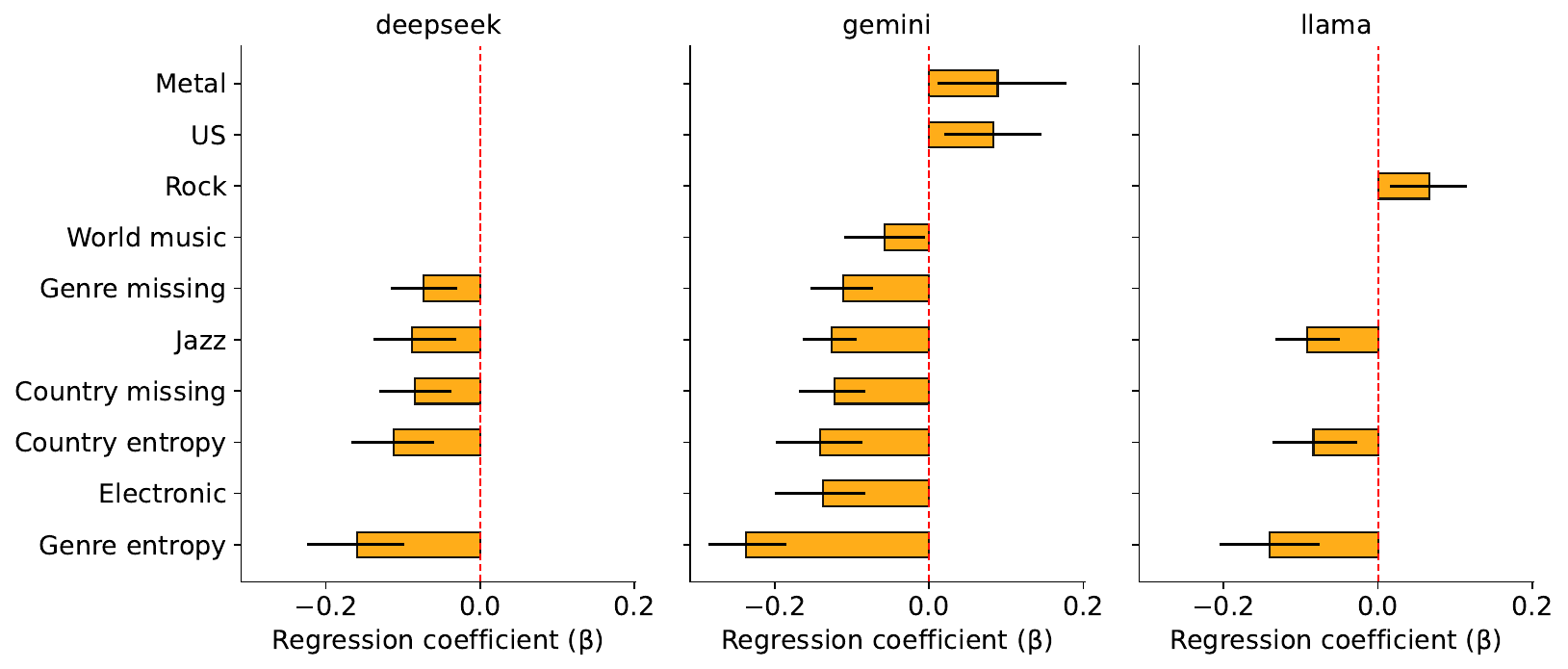}
    \caption{5K bootstrapped estimated linear coefficients of sampled item characteristics predicting recall@10.}
    \label{fig:item_recall}
\end{figure*}

\begin{figure*}
    \centering
    \includegraphics[width=0.8\linewidth]{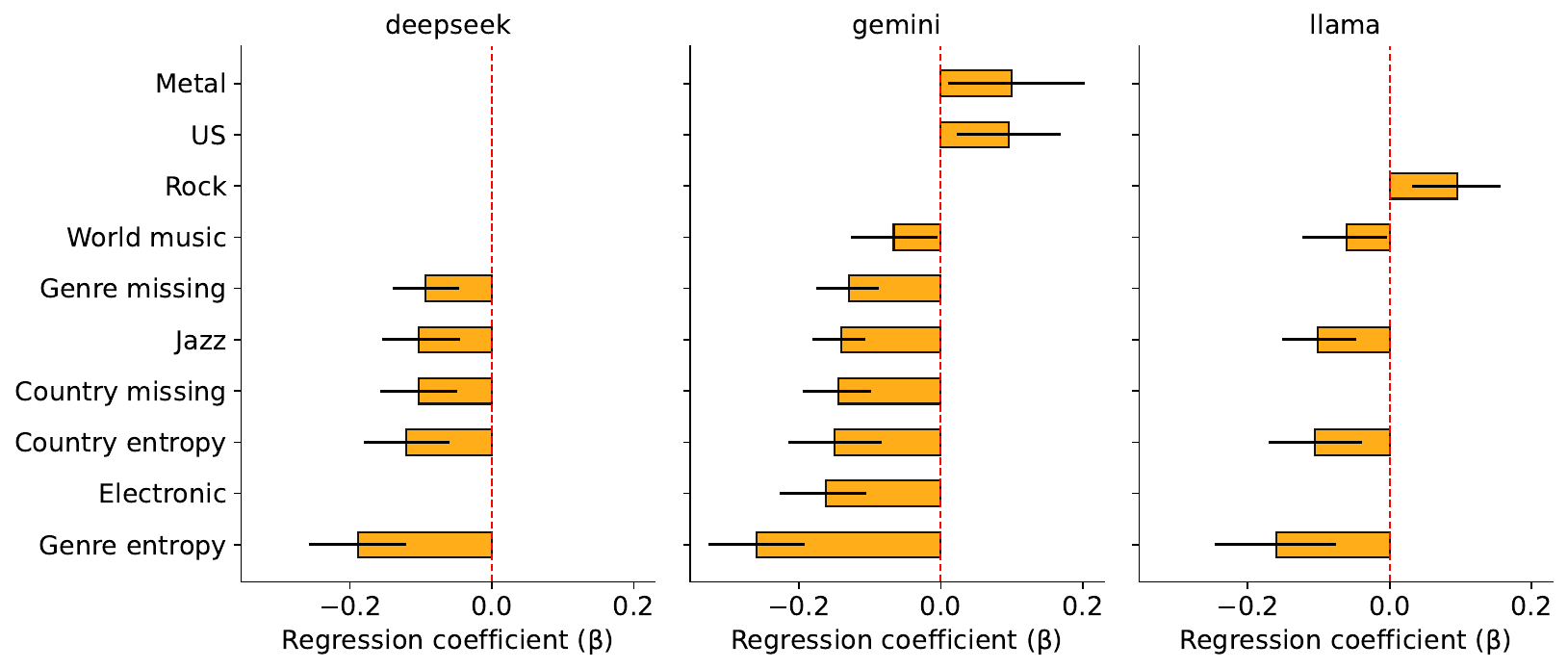}
    \caption{5K bootstrapped estimated linear coefficients of sampled item characteristics predicting ndcg@10.}
    \label{fig:item_ndcg}
\end{figure*}

Figures~\ref{fig:item_recall} and~\ref{fig:item_ndcg} present the bootstrapped mean $\beta$ coefficients and 95\% CI from 5K linear regressions predicting Recall@10 and NDCG@10 based on the sampled item characteristics. We exclude associations with CI intervals crossing zero or with width greater than 10, as these indicate high uncertainty and lack of a reliable effect.

\section{Full Prompt and NL profile}

For generating the profiles we used the same prompt format with a different list of items. All prompts started with the same preamble, followed by the item metadata and finishing with ``Only return the textual summary — no explanation, no lists.'' We do not include profile examples in the prompt, allowing for flexibility. 

In this section, we show different model generations given the following input prompt, with items collected from a 90 day time-window: 

\noindent\rule{\linewidth}{1pt}

\textit{Here is a list of items consumed by a single user. From this list, generate a short and scrutable summary of their musical taste for music recommendation purposes. The summary should be a single, coherent, and informative paragraph that clearly conveys the user's preferences in a natural, human-readable way.  \\  \\  Base the description on the following list of consumed items and their metadata, without focusing on the names of tracks and artists. The items are presented in descending order: \\  \\ title: This is a Mood\\ playcount: 19\\ artist: Franc Moody\\ origin country: United Kingdom\\ album release date: 2020\\ main genre: Electronic music\\ secondary genre: Nu-disco, Synthpop, Electronica\\ \\ title: CAFé CON RON\\ playcount: 11\\ artist: Bad Bunny\\ origin country: Puerto Rico\\ album release date: 2025\\ main genre: Rap music\\ secondary genre: Caribbean music, Dancehall, Dance rap, Latin hip hop, Latin\\ \\ title: Te Felicito (Turreo Edit)\\ playcount: 9\\ artist: Shakira\\ origin country: Colombia\\ album release date: 2023\\ main genre: Pop music\\ secondary genre: Latin trap, Pop rock, Electro pop, Contemporary R\&B, Reggaeton, Latin american music\\ \\ title: Y sólo se me ocurre amarte (Unplugged)\\ playcount: 9\\ artist: Alejandro Sanz\\ origin country: Spain\\ album release date: 2004\\ main genre: Pop music\\ secondary genre: Pop rock, Latin\\ \\ title: Feel So Close (As Made Famous By Calvin Harris Instrumental Cover)\\ playcount: 9\\ artist: Calvin Harris\\ origin country: United Kingdom\\ album release date: 2012\\ main genre: Electronic music\\ secondary genre: Electro pop, Nu-disco, Urban pop music, Electronic dance music (EDM), Contemporary R\&B\\ \\ title: Dog Music (Sleeping Music)\\ playcount: 8\\ artist: None\\ origin country: None\\ album release date: 2017\\ main genre: None\\ secondary genre: None\\ \\ title: Llamado De Emergencia Deluxe\\ playcount: 6\\ artist: Daddy Yankee\\ origin country: Puerto Rico\\ album release date: 2024\\ main genre: World music\\ secondary genre: Latin electronic, Dembow, Dance pop, Latin pop music, Dancehall, Latin hip hop, Latin urban, Latin, Latin american music\\ \\ title: Entra En Mi Vida\\ playcount: 5\\ artist: Sin Bandera\\ origin country: Mexico\\ album release date: 2001\\ main genre: Rock music\\ secondary genre: Latin pop music\\ \\ title: Big Zen\\ playcount: 4\\ artist: None\\ origin country: None\\ album release date: 2023\\ main genre: None\\ secondary genre: None\\ \\ title: La Travesia\\ playcount: 4\\ artist: Juan Luis Guerra\\ origin country: Dominican Republic\\ album release date: 2007\\ main genre: World music\\ secondary genre: Reggae music, Salsa, Latin pop music, Merengue music, Latin, Cumbia, Bolero\\ \\ title: 396 Hz Dissolve Negativity\\ playcount: 4\\ artist: None\\ origin country: None\\ album release date: 2019\\ main genre: None\\ secondary genre: None\\ \\ title: Beautiful Liar\\ playcount: 3\\ artist: Beyoncé\\ origin country: United States of America\\ album release date: 2007\\ main genre: Soul music\\ secondary genre: Pop rap, Soul pop, Pop rock, Contemporary country\\ \\ title: 396\\ playcount: 3\\ artist: Timo Krantz\\ origin country: United States of America\\ album release date: 2020\\ main genre: Electronic music\\ secondary genre: None\\ \\ title: Never Had A Dream Come True\\ playcount: 2\\ artist: S Club\\ origin country: United Kingdom\\ album release date: 2000\\ main genre: Pop music\\ secondary genre: Contemporary R\&B, Television series\\ \\ title: Abreme Paso\\ playcount: 2\\ artist: None\\ origin country: None\\ album release date: 2021\\ main genre: None\\ secondary genre: None\\  \\  Only return the textual summary — no explanation, no lists.}

\noindent\rule{\linewidth}{1pt} 

From the prompt above, the following profiles were generated:
\vspace{10pt}

\begin{itemize}
    \item \textbf{Deepseek-r1} (user rating: 4.0.): \textit{The user has a strong preference for electronic music, particularly in genres like nu-disco, synthpop, and EDM, supported by notable tracks such as "Mood" and "Feel So Close." They also enjoy pop music with elements of Latin influence, as seen in works by Shakira and others. Additionally, their tastes include rap and rock, showcasing a diverse yet cohesive musical style that blends electronic, pop, and Latin flavors.}
\vspace{10pt}
    \item \textbf{Gemini-2.0-flash} (user rating: 7.0.): \textit{This user enjoys a diverse mix of music, with a strong preference for electronic music, particularly nu-disco and synthpop from the United Kingdom. They also have a significant interest in Latin music, including Latin pop, reggaeton, and Latin hip hop from various Latin American countries and Puerto Rico, alongside some mainstream pop and occasional soul and rock influences. They also seem to use music for relaxation or background ambiance, as indicated by tracks explicitly intended for sleep or relaxation.}
\vspace{10pt}
    \item \textbf{Llama3.2:3b} (user rating: 4.0.): \textit{This user's musical taste leans towards a blend of electronic and pop elements, with a strong appreciation for global influences, particularly from Latin America and the Caribbean. They seem to enjoy upbeat and energetic tracks, often incorporating tropical rhythms and melodies, as evidenced by their fondness for artists like Bad Bunny, Shakira, and Daddy Yankee. There's also a noticeable interest in calming and atmospheric soundscapes, such as the instrumental cover of "Feel So Close" and the ambient piece "Dog Music". This user appears to have a diverse musical palette, but their overall taste tends towards lively and danceable tracks with Latin flair.}

\end{itemize}

\section{User feedback}

Some users left us feedback on their profiles, here we list the more informative ones. Users were overall happy with their profiles, some complained about hallucinations, others about missing parts of their tastes due to the sampled period.
\vspace{1em}

\begin{quote}
``The AI definitions can be long and repetitive at times but the exercise is interesting.''
\end{quote}

\vspace{1em}

\begin{quote}
``Some artists were picked as an example of a sub-genre I'm supposed to listen to, but sometimes the artist was misplaced (for example: Whomadewho has a rock band instead of electronic), leading to a wrong description of my music tastes.''
\end{quote}

\vspace{1em}

\begin{quote}
``I would love to compare what I declared against what I actually do. But in general, loved being part of it!!! Made me reflect a lot on how I value things I listen in a more lean in way.''
\end{quote}

\vspace{1em}

\begin{quote}
``Nice work, there were some options that were totally relevant. 
The ones that I preferred were those with the name of artists I listened too. Some descriptions were including the country origin of artists and it seemed like a hit or miss for me. As I don't know the origin of all artists I listen too, there were some descriptions where I wasn't sure if they were adequate.''
\end{quote}

\vspace{1em}

\begin{quote}
``Seems to be missing quite a big part of my taste in music so I'm wondering if the period used for this test is sufficient (I listen to a lot of Reggae usually and it hasn't surface at all in the description - but if the sample used for test is too short, it might not include reggae indeed).''
\end{quote}

\vspace{1em}

\begin{quote}
``Many of the profiles presented hallucinations with track names that don't exist or were misattributed to the wrong artist or music genre. The longer and more detailed the description tried to be, the more these hallucinations appeared.''
\end{quote}

\vspace{1em}

\begin{quote}
``20\% of times, you seems to have picked an outliers genres, that I don't care about at all, but that matches one specific track or artist that I streamed a lot to. Like track ``4 Kampé'' (Zouk Love) or artist ``Daft Punk'' (Electro). Also, very sad to not see genre that I actually love (Carioca Funk, Afro Beat) but don't stream very often.
Mainly because I don't know how to find good tracks for these genres.
It was a fun ride nonetheless. Thank you!''
\end{quote}

\end{document}